\documentclass[letterpaper,12pt,preprint]{aastex}
\usepackage{enumitem, color}
\definecolor{hypercolor}{RGB}{0,0,127}
\usepackage[%
  citecolor=hypercolor,%
  linkcolor=hypercolor,%
  urlcolor=hypercolor,%
  backref=false,%
  pagebackref=false%
]{hyperref}%

\newcommand{\sectionname}{Section}
\newcommand{\documentname}{\textsl{white paper}}
\newcommand{\foreign}[1]{\textit{#1}}

\newcommand{\etal}{\foreign{et~al.}}
\newcommand{\observatory}[1]{\textsl{#1}}
\newcommand{\Kepler}{\observatory{Kepler}}

\newcommand{\SDSS}{\observatory{SDSS}}
\newcommand{\WISE}{\observatory{WISE}}
\newcommand{\project}[1]{\textsl{#1}}
\newcommand{\MAST}{\project{MAST}}
\newcommand{\kplr}{\project{kplr}}
\newcommand{\TheTractor}{\project{The~Tractor}}
\newcommand{\emcee}{\project{emcee}}
\newcounter{inlineitem}
\newcommand{\binlineitem}{\refstepcounter{inlineitem}\textbf{\textsl{(\theinlineitem)}}}
\newcommand{\inlineitem}{\refstepcounter{inlineitem}{\textsl{(\theinlineitem)}}}
\newcounter{address}
\newcommand\independent{\protect\mathpalette{\protect\independenT}{\perp}}
\def\independenT#1#2{\mathrel{\rlap{$#1#2$}\mkern2mu{#1#2}}}

\begin{document}\sloppy\sloppypar\thispagestyle{empty}

\title{Maximizing \Kepler\ science return per telemetered pixel: \\
  Detailed models of the focal plane in the two-wheel era\altaffilmark{\ref{kcall}}}

\author{%
  David~W.~Hogg\altaffilmark{\ref{CCPP},\ref{MPIA},\ref{email}},
  Ruth~Angus\altaffilmark{\ref{Oxford}},
  Tom~Barclay\altaffilmark{\ref{Ames}},
  Rebekah~Dawson\altaffilmark{\ref{CfA}},
  Rob~Fergus\altaffilmark{\ref{Courant}},
  Dan~Foreman-Mackey\altaffilmark{\ref{CCPP},\ref{datadriven}},
  Stefan Harmeling\altaffilmark{\ref{MPIIS}},
  Michael~Hirsch\altaffilmark{\ref{UCL},\ref{MPIIS}},
  Dustin~Lang\altaffilmark{\ref{CMU},\ref{physical}},
  Benjamin~T.~Montet\altaffilmark{\ref{Caltech}},
  David~Schiminovich\altaffilmark{\ref{Columbia}},
  Bernhard~Sch\"olkopf\altaffilmark{\ref{MPIIS}}%
}

\setcounter{address}{1}
\altaffiltext{\theaddress}{\stepcounter{address}\label{kcall}%
  A \documentname\ submitted in response to the \Kepler\ Project Office
  \textit{Call for White Papers: Soliciting Community Input for
    Alternate Science Investigations for the Kepler Spacecraft}
  released 2013 August 02
  (\url{http://keplergo.arc.nasa.gov/docs/Kepler-2wheels-call-1.pdf})}
\altaffiltext{\theaddress}{\stepcounter{address}\label{CCPP}%
  Center for Cosmology and Particle Physics, Department of Physics, New York University}
\altaffiltext{\theaddress}{\stepcounter{address}\label{MPIA}%
  Max-Planck-Institut f\"ur Astronomie, Heidelberg, Germany}
\altaffiltext{\theaddress}{\stepcounter{address}\label{email}%
  To whom correspondence should be addressed; \texttt{<david.hogg@nyu.edu>}.}
\altaffiltext{\theaddress}{\stepcounter{address}\label{Oxford}%
  Department of Physics, Oxford University}
\altaffiltext{\theaddress}{\stepcounter{address}\label{Ames}%
  NASA Ames Research Center}
\altaffiltext{\theaddress}{\stepcounter{address}\label{CfA}%
  Harvard--Smithsonian Center for Astrophysics}
\altaffiltext{\theaddress}{\stepcounter{address}\label{Courant}%
  Courant Institute of Mathematical Sciences, New York University}
\altaffiltext{\theaddress}{\stepcounter{address}\label{MPIIS}%
  Max-Planck-Institut f\"ur Intelligente Systeme, T\"ubingen}
\altaffiltext{\theaddress}{\stepcounter{address}\label{UCL}%
  Department of Physics and Astronomy, University College London}
\altaffiltext{\theaddress}{\stepcounter{address}\label{CMU}%
  McWilliams Center for Cosmology, Carnegie Mellon University}
\altaffiltext{\theaddress}{\stepcounter{address}\label{Caltech}%
  Department of Astronomy, California Institute of Technology}
\altaffiltext{\theaddress}{\stepcounter{address}\label{Columbia}%
  Department of Astronomy, Columbia University}
\altaffiltext{\theaddress}{\stepcounter{address}\label{datadriven}%
  DFM is responsible for most of the data-driven modeling shown in this \documentname.}
\altaffiltext{\theaddress}{\stepcounter{address}\label{physical}%
  DL is responsible for most of the toy data simulations and the physical modeling shown in this \documentname.}

~
\clearpage

\section{Executive summary}

\binlineitem~The \textbf{primary recommendation} of this \documentname\
  is \emph{image modeling}, that is,
  fitting to the pixel values downlinked from the \Kepler\ satellite
  a quantitative description of pixel sensitivities (flat-field),
  point-spread function (PSF),
  and sky, bias, and dark signals,
  all as a function of focal-plane position.
This kind of modeling has not been required for \Kepler\ so far
  because its aperture-photometry precision has been maintained through pointing stability.
The degraded pointing precision in two-wheel operations will be a blessing as well as a curse:
It reduces the precision of aperture photometry,
  but it provides diversity that permits inference
  of the detector sensitivity map and PSF.
\textbf{We propose developing a probabilistic generative model of the \Kepler\ pixels.}
We argue that this may permit continuance of photometry at 10-ppm-level precision.
We demonstrate some baby steps towards precise models
  along both data-driven (flexible) and physics-driven (interpretably parameterized)
  directions.
We demonstrate that the expected drift or jitter in positions in the two-weel era
  will help enormously with constraining calibration parameters.
In particular, we show that we can infer the device flat-field at \emph{higher than pixel resolution};
  that is, we can infer pixel-to-pixel variations in intra-pixel sensitivity.
These results are relevant to \emph{almost any} scientific goal for the repurposed mission;
  image modeling ought to play a role no matter what.
We have several secondary recommendations:

\binlineitem~It will be imperative to operate the spacecraft
  with modifications to tables or software:
Either the pointing will need to be adjusted frequently using
  the two operational wheels and some propellant;
  or else the telemetered focal-plane apertures will have to be enlarged;
  or else the apertures will have to be adapted in real time to follow drifting stars.

\binlineitem~It will be wise to perform
  deliberate focus pulls, dithers, and integration-time adjustments;
  these will provide much more data support for hard-to-constrain calibration parameters.
These calibration observations will improve any two-wheel data but will also
  improve the precision
  of the extant \Kepler\ data that we expect contains many undiscovered signals.

\binlineitem~It is our view that \Kepler\ ought to continue work on
  what many of the present authors consider its key scientific goal,
  which is to continue to \textbf{find Earth-like planets on year-ish orbits around Sun-like stars}.
With a multi-pronged image-modeling effort and
  a bit of good luck
  (with respect to assumptions about spacecraft hardware and data-modeling software),
  it is our view that in the next few years \Kepler\ can do even more
  than it already has on this deep and important mission.
That said, what's written in this \documentname\ is fundamentally
  agnostic about the scientific program in the two-wheel era.

\clearpage
\section{Philosophy and motivation}

The \Kepler\ satellite%
  ---like other very precise time-domain photometric projects---%
  achieves its precision by keeping the camera and telescope exquisitely controlled and stable,
  leading to identical conditions for each observation.
This strategy has yielded large numbers of planets in a wide range of sizes and periods.

In the two-wheel era, it will not be possible
  to make each observation as identical to the fiducial observation
  as it was in the three- and four-wheel eras.%
  \footnote{See, for example, the
  \textit{Call for White Papers: Soliciting Community Input for Alternate Science Investigations for the Kepler Spacecraft}
  (\url{http://keplergo.arc.nasa.gov/docs/Kepler-2wheels-call-1.pdf})
  and the
  \textit{Explanatory Appendix to the Kepler Project call for white papers:\ Kepler 2-Wheel Pointing Control}
  (\url{http://keplergo.arc.nasa.gov/docs/Kepler-2-Wheel-pointing-performance.pdf}).}
The imprecise pointing introduces, into any description of the data, three unknown functions of time
  (that is, the three solid-body Euler angles).
These three Euler-angle functions affect—--in a predictable way---%
  every single one of the hundreds of thousands of detectable stars
  in the camera field of view,
  each of which affects the intensity field touching a few to many pixels.
Therefore the photometry can be used to constrain these three functions,
  which are sampled at different value during each exposure;
  indeed, there is an \emph{overwhelming abundance} of information in the data
  to constrain these three functions.

While it is obvious \emph{in principle} that we can infer
  the intra-exposure s/c pointing and orientation history,%
  \footnote{Indeed inference of non-trivial intra-exposure
    camera pointing trajectories has been demonstrated in the far more difficult
    case of single-exposure ``natural'' images
    (meaning ordinary personal photographs of everyday scenes; \citealt{fergus2006,whyte2010,hirsch2011,koehler2012}).
  That is, the three-axis history of camera orientation path
    can be reliably inferred in an individual image,
    without even multiple exposures to use
    and no knowledge of the details of the ``true scene''.
  Relative to these situations, the \Kepler\ problem is a walk in the park!
  That said, the precision requirements of \Kepler\ science
    are much more challenging than those of natural image de-blurring.}
  and thus potentially brightness variations of stars limited only by photon noise,
  we cannot perform a convincing test with the existing \Kepler\ data.
Recovery of precise photometry from a source
  that is moving across the focal plane
  requires knowledge of not just the pointing but the device geometry and sensitivity
  (flat-field) at the pixel level (or probably at even \emph{higher} resolution),
  along with the point-spread function (PSF) and its variations with time and position.
When---as with \Kepler---all data have been taken at rigidly fixed pointing,
  it is hard (or maybe impossible) to obtain independent data-supported models of
  the PSF and the flat-field,
  let alone any intra-pixel sensitivity variations
  or pixel-level additive signals (like bias or dark currents).
That is,
  \Kepler\ data are taken in a mode that is
  near-pessimal%
  \footnote{Neologism ``pessimal'' is the antonym of ``optimal''.}
  for the determination of calibration parameters.

In the demonstrations that follow,
  we will show that there is great promise for recovery or determination of precise calibration information
  in the two-wheel era.
This \documentname\ does not deliver an air-tight proposal for obtaining 10-ppm photometric precision.
It attempts only to \emph{point the way} towards extremely precise modeling,
  something which should be a part of any future scientific program with the
  degraded satellite.
The key idea is that the drift of the stars across the CCD provides
  \emph{precisely the data we need} to determine the calibration functions.
A multi-pronged image modeling effort along the lines outlined here
  is almost certain to deliver large information gains to any future (or even past) \Kepler\ science program
  at comparatively low cost.

The death of \Kepler's second wheel (taking it down to two)
  occurred at a tragic moment:
The mission is on track to discover Earth-like planets
  orbiting Sun-like stars on year-ish (habitable-zone) orbits,
  and for such targets, five or six years is a lot better than four.
Indeed, when signals are sparse, statistical significance
  (as estimated by, say, a p-value)
  can grow much faster than any small power of mission lifetime.%
  \footnote{We thank Ben Weiner (Arizona) for this insight.}
It is our view that discovery and characterization of exoplanets is so important,
  it should remain \Kepler's top priority.
Additional data collected with two-wheel \Kepler\ could
  significantly enhance the number and fidelity of Earth-analog-candidates
  discovered by \Kepler\ (and
  also be extremely valuable for \Kepler's other science goals).
With this in mind, we advocate using \emph{methods of image modeling} to retain
  or approximate three-wheel photometric precision
  in an era of two-wheel pointing degradation.
If that is possible---and we think it will be---we would recommend continuing work
  along the direction of \Kepler's core mission,
  which (in our view) is to continue finding and characterizing
  Earth analogs and other habitable-zone oceany rocks.

Of course the two-wheel mode will bring other challenges, like
  possible hard constraints on field choice,
  changes to exposure time or cadence,
  and changes to flight tables or software that selects pixel regions for downlink.
For all these reasons, we might have to observe fewer stars in the two-wheel era.
Since \Kepler's mission so far has taught us a lot about the variability
  properties of stars,
  the prevalence of rocky planets around different kinds of stars,
  and the brightness limits of transit searching at different transit depths and durations,
  it is our view that it will be possible to design a program that will deliver
  large payloads of new and important planetary systems,
  and all relevant to the Earth-analog question.
We present some of these scientific proposals in a separate, companion white paper (Montet \etal),
  because although we specifically happen to love eta-Earth,
  \emph{what we say here about image modeling is equally applicable
  to any two-wheel repurpose} for the \Kepler\ satellite.

Right now, all of \Kepler's spine-tingling precision
  has been obtained by peforming rigid integer-pixel aperture photometry
  on flat-fielded and background-subtracted image data.
Furthermore, only something like half of the downlinked pixels are used in the apertures,
  and every pixel that \emph{is} used is used with identical weight.
That is, there is no ``optimal extraction'' or ``PSF photometry'' being
  used to combine the corrected pixel data;
  the PSF model is used only to determine optimized photometric apertures.
Another motivation for and value of the work we present here will be
  in the exploitation of the \emph{existing} data.
In some sense, one of the proposals we are making is to use the two-wheel era
  to generate the calibration data we need to fully exploit the existing data.
Again, this calibration activity could bring enormous scientific value to the \Kepler\ community
  no matter what the repurposed mission.

One final point of philosophy:
All of the code used to make the demonstrations in this \documentname\ is available
  in open-source or at least publicly readable repositories on the web.%
\footnote{Our toy future-data simulation---and code for physical modeling thereof---is at
  \url{http://astrometry.net/svn/trunk/projects/kepler/};
  it is a sub-project of our general image-modeling framework
  called ``\TheTractor'' (\citealt{hoggtractor}, \citealt{langtractor}), which is at
  \url{http://theTractor.org/}.
  Our data-driven modeling code is at
  \url{https://github.com/dfm/causal-kepler}.
  Code to build physical models of the pixels in the extant \Kepler\ data is at
  \url{https://github.com/dfm/kpsf}.
  Our Python wrapping of the \MAST\ interface to the \Kepler\ data is at
  \url{https://github.com/dfm/kplr},
  and most of our probabilistic inference involves ensemble sampling
  with \emcee\ (\citealt{emcee}), which is at
  \url{http://dan.iel.fm/emcee}.
  This \documentname\ itself was developed and lives at
  \url{https://github.com/davidwhogg/SaveKepler}.}
We encourage the \Kepler\ community to fork, build on, and contribute back to these code bases.

\section{Data-driven modeling}\label{sec:datadriven}

In this \documentname, we sketch several methods for modeling both existing
\Kepler\ data and simulated 2-wheel observations.
The discussion here should be viewed as an illustration of potentially
interesting data analysis techniques and not as a definitive (or even
fully-formed) argument.
That being said, the preliminary results seem promising.
The methods that we explore fall into two main categories: data-driven models
and physically motivated generative models.
We will start by describing some data-driven methods in this section and then
demonstrate some toy examples of generative modeling in the subsequent
sections.
In this section, we draw on expertise in the fields of machine learning,
causal inference and computer vision to propose non-parametric models for systematics removal in \emph{existing} \Kepler\ observations.
Most 2-wheeled observing strategies will amplify existing systematics.
With this in mind, it is clear that improving the precision of techniques used
to analyze existing data will likely be applicable to the data taken in
suboptimal operational modes.
Some of the basic ideas in our discussion of data-driven modeling should sound
familiar to \Kepler\ experts because the PDC techniques employed by the core
data reduction pipeline are excellent examples of this type of model.

\subsection{Predicting pixels with pixels}\label{sec:pixels}
The PDC algorithms \citep{map-pdc1,map-pdc2} are particularly
successful because they remove instrumental effects (assumed common between
nearby targets on the detector) while retaining astrophysical signals (assumed
unique to the target of interest).
These methods use a data-driven linear model built directly from the aperture
photometry of nearby stars; that is, they operate on coadded pixel values,
not individual pixels.

Here, we present a demonstration in the form of an extension to this model
that explicitly takes advantage of the causal structure of the problem (CITE
SOMETHING).
We also argue that this procedure is best performed \emph{at the pixel level}
instead of on the extracted photometric time series.
To start, we describe a model to remove systematic effects from the time
series in a single pixel using the fluxes of pixels from some $K\ge1$ nearby
star(s) and the ``historical'' time series of the target pixel.
The main insight of this method is that the systematic effects are actually
predictable based on the history of the system and not just on the current
state.
In other words, the (recent) history (and possibly the future) of the detector
sensitivity, as encoded in large numbers of pixel histories, provides a
better estimate of the conditions at a given moment than an
instantaneous measurement considered alone.

In this section,
we will discuss the calibration of pixel $i$ of target star $n$ at time $t_k$
using the pixels $i^\prime$ on target $m \ne n$ in the sliding window $t_k -
\delta t \le t_{k^\prime} \le t_k + \delta t$.

We start by making the (reasonable) assumption that the observed signal---the
flux measured in pixel $i$ at time $t_k$: $y_i (t_k)$---is a function of only
the ``true'' flux $f_i(t_k)$ arriving at the location of pixel $i$ during the
integration time,
bulk variations caused by the conditions of the spacecraft (for example,
pointing shifts and temperature fluctuations) that affect every pixel in a
systematic---but not identical---way, and
instrumental noise due to the detector.
The goal is to find (or model) the true fluxes $f_i(t_k)$ conditioned on the
observations $y_i(t_k)$.
In practice, this is hard because we don't have a physical model accurate at
the required levels.
Instead, we propose an effective model based on the observations described
above.

Our method relies on assumptions of ``independence'' and ``join confounding''.
\emph{Independence} (motivated by Reichenbach's principle; \citealt{Reichenbach1956}):
we assume that the true fluxes are independent: $f_i(t_k) \independent
f_{i^\prime} (t_{k^\prime})$ for all $i \ne i^\prime$ 
We also assume there may be additional observables (for example, temperature)
for which this independence holds.
It is important to note here that we are only considering pixels $i^\prime$
from a \emph{different} star than the star targeted by the pixel $i$.
\emph{Joint confounding}:
we assume that any confounding effect on $y_i(t_k)$ will also affect a number
of other values $y_{i^\prime} (t_{k^\prime})$ (where $i \ne i^\prime$ and
$t_{k^\prime}$ is not necessarily equal to $t_k$).
This means that the overall systematic affecting $y_i(t_j)$ can (in
principle) be predicted from from the other $y_{i^\prime}(t_{k^\prime})$.

The idea behind our approach is that if we now try to predict a measurement
$y_i (t_k)$ from variables known to be \emph{independent} of $f_i (t_k)$ (the
true flux), then whatever we \emph{can} predict has nothing to do with
$f_i (t_k)$.
The statistical dependence
we find must, by Reichenbach's principle, be due to joint confounders that
affect both our measurement and the other variables we are using to predict
it.

For simplicity, we assume that the systematic confounders affect the signal
additively (although it may be possible to generalize the method).
Under this assumption, the model can be written as
\begin{eqnarray}\label{eq:reg-model}
y_i(t_k) &=& f_i (t_k) + \sum_{i^\prime,\,k^\prime} c_{i^\prime\,k^\prime}\,
    y_{i^\prime} (t_{k^\prime}) \quad.
\end{eqnarray}
Since (by our independence assumption) $f_i(t_k) \independent
y_{i^\prime} (t_{k^\prime})$, the vector of coefficients $\mathbf{c}$ can be
computed using using standard linear least-squares techniques on the system
in equation~\ref{eq:reg-model}, treating $f_i (t_k)$ as the noise term.

\begin{figure}
\includegraphics[width=0.9\textwidth]{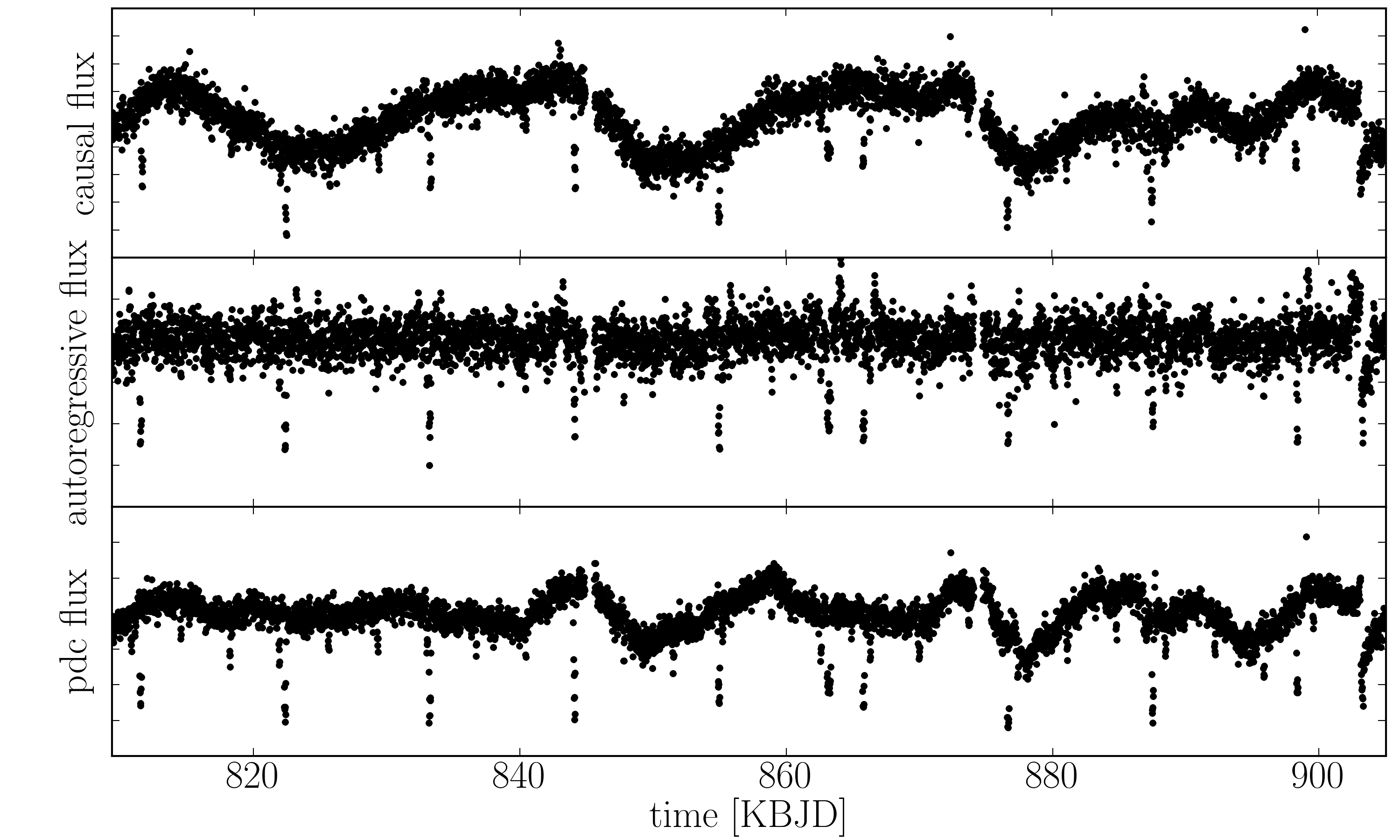}%
\caption{Results of our data-driven pixel-level systematics model applied to
the quarter 9 observations of \Kepler-20 and compared to the PDC light curve.
\textsl{(top)}~The basic model (equation~\ref{eq:reg-model}) applied to the
pixel time series. This \figurename\ shows the results of coadding the
\emph{residuals} ($f_i (t_k)$ in equation~\ref{eq:reg-model}) using the
optimal aperture from the \Kepler\ pipeline.
\textsl{(middle)}~The same as the top panel using an autoregressive model
(with $\Delta = 12\,\mathrm{hours}$).
\textsl{(bottom)}~The results of doing simple aperture photometry and then
running PDC on the extracted light curve. \label{fig:reg-model}}
\end{figure}

To demonstrate this method, we applied equation~\ref{eq:reg-model} to the
quarter 9 observations of \Kepler-20, a G-type star with 5 known transiting
exoplanets \citep{kepler20}.
The results are shown in the top panel of \figurename~\ref{fig:reg-model}.
To generate this \figurename, we downloaded the target pixel files from
\MAST\footnote{The Mikulski Archive for Space Telescopes;
\url{http://archive.stsci.edu/}} using
\kplr\footnote{\url{http://dan.iel.fm/kplr}}
and computed the vector $\mathbf{c}$ for each pixel using the pixel time
series from the two nearest stars on the detector as the basis.
The result shown in the top panel of \figurename~\ref{fig:reg-model} is the
result of coadding the time series using the optimal aperture provided by the
\Kepler\ photometry module.
It is instructive to compare this to the PDC result (shown in the bottom panel
of \figurename~\ref{fig:reg-model}) to see that we achieve similar stellar
variability signatures and preserve the transit signal.
We have not yet performed any quantitative comparison of these two methods but
the fact that we were able to achieve qualitatively competitive performance
without exploiting the full flexibility of this kind of model is very promising.

\subsection{Autoregressive generalization}
The model discussed in the previous section should retain \emph{any true
astrophysical signals} because there is no physical mechanism that would
induce correlations between these signals in different targets.
Suppose that we are only interested in temporally compact astrophysical
signals---exoplanet transits, for example.
In this case, we can substantially improve the predictive power of the model
by using information from the (recent) history of the pixels targeting the star
that we are trying to model.
That is, by using not just pixels from other stars, but using pixels
from (an appropriately chosen) time-window in past history of the
current star, we can model not just the instrumental variations but
also the stellar variations, without messing up the transits.

In this case, observations of the target of interest are independent of observations of
\emph{other stars} for times $t_{k^\prime}$, where $|t_k - t_{k^\prime}| >
\Delta$ and $\Delta$ is longer than the maximum duration of a transit.
This independence assumption has a physical interpretation: in addition
to all the above independences, the flux of a planet-hosting star is
independent of planet position \citep[except in rare cases of close-in planets
that produce phase variations,][]{esteves2013}. Therefore, the light arriving
from the star system contains no information about the transit signal except
while the planet is in front of (or behind) the star.
This can be viewed as an assumption of independence of the stellar limb
darkening profile and the geometric mechanism by which the
star-planet system ``converts'' the star brightness into an observable
brightness of the joint system.
This kind of assumption has recently studied in the field of causal inference
(\citealt{JanSch10}).

To demonstrate this method, we preformed an autoregressive fit to the same
data as discussed in the previous section (quarter 9; \Kepler-20).
The results of this fit are shown in the middle panel of
\figurename~\ref{fig:reg-model}.
The difference between this and the previous demonstration is that, in this
case, a 2.5 hour window is included from the pixels in the \Kepler-20 aperture
$\Delta = 12\,\mathrm{hours}$ before and after the target time are included in
the basis.
This model removes most of the stellar variability in the light curve
  \emph{without significantly reducing the transit signal.}
Again, we do not yet have a sound quantitative argument demonstrating the power of
this method but a robust detrending algorithm applied at the pixel level could
be extremely valuable and these results show promise.

\subsection{Deep learning}\label{sec:deep}

In contrast to the linear conditional models outlined above, we also
recommend exploring a range of highly non-linear models for the same
task. While many of the confounding signals will be linear-ish,
the residuals will contain non-linear
factors in the system that might obscure the transit signal. To leave
only the transit signal in the residual, these non-linear but unknown factors much
be captured with a more powerful model. This can be done using \emph{Deep
Learning} approaches that have recently proven very successful for
complex image domain tasks such as object recognition \citep{Kriz12}.

The Deep Learning models we intend to apply in this instance will be a
convolutional neural network (CNN), similar in spirit to the LeNet
architecture (\citealt{LeCun1998}). This model is a
multi-layered (hence ``deep'') and has special connectivity structure so that
each unit in a layer examines only a small window of the previous one
(rather than all units). Each layer computes a convolution of the
feature map from the previous layer (a 2D map) with a learned filter
and then the result is passed through a non-linear function.  Models with many layers can capture
highly non-linear structure of the input image, since each unit in a
higher layer is connected to the input by multiple non-linear
functions.  We recommend three approaches:

\setcounter{inlineitem}{0}\inlineitem~\emph{Spatial-only}: In this model, we just attempt to model the spatial
structure within a single 2D image. The input will be a set of stars
around a pixel whose brightness we wish to predict (the immediately
surrounding pixels will be masked). The output will be the brightness
of the aforementioned pixel.  When trained
on a set of 2D images, taken at different times, the early layers of
the model might be expected to learn the PSF, while the higher layers
will learn the dependencies between the brightnesses of stars (which
should just be from confounding sources).

\inlineitem~\emph{Temporal-only}: The drawback to the spatial-only model is that it
cannot capture the temporal dependencies in the data.  Instead, we
can take as input time series from pixels (centers of nearby
stars, say) and the time series of the star of interest as output. The
LeNet model architecture uses 2D convolutions, applied to a 2D input
image. This must be modified to have a 1-D convolution in the time
direction, but fully connected across the pixels. The size of the
convolutional window dictates the amount of ``history'' the model can
use for prediction.

\inlineitem~\emph{Spatio-temporal}: This combines the above models and allows spatial
and temporal structure to be learned. The input is a 3D sequence of
images, with 3D convolutional filters used to produce 3D feature maps.
The output will be sequence of patches, containing the star of
interest over time.

All three models will use data split into three disjoint sets for
\emph{training, validation and testing}, all screened to avoid any
transits where possible. Selecting the appropriate model architecture (number of
layers, number of feature maps per layer) is an unsolved problem in
machine learning, thus can only be done by training a range of
different model variants and selecting the best using the validation
set. If the model is too large it will have too much capacity, thus
overfit the training set. Conversely, a small model will not have
sufficient power to predict the target star with enough fidelity to
reveal the transit.

\section{Physical modeling}\label{sec:physical}

Data-driven models of the types described above can work
  because the relationships between the different data flowing from the system
  (telemetered pixel values, in this case)
  are determined by an underlying set of physical devices
  (s/c, optics, detectors, and electronics, in this case),
  each of which is deterministic or stationary or at least predictable probabilistically.
Data-driven models have the advantage of flexibility%
  ---we don't have to decide in advance exactly what freedoms to give the model---%
  but they have the disadvantage that they don't represent all of our prior beliefs
  about what \emph{can} happen and why.
For these reasons, it is worth thinking about physical models.
In a physical model, we build highly parameterized models
  of the s/c attitude, PSF, flat-field (and possibly intrapixel sensitivity),
  and DC signals.
These are compared with the data probabilistically,
  through a likelihood function.
For the purposes of demonstrating the physical modeling direction,
  in this part of the \documentname\, we use simulated \Kepler\ data
  to demonstrate feasibility.
As with the previous \sectionname, the demonstrations and discussions here are designed
  to be \emph{illustrative};
  we have not yet proven that 10~ppm precision is straightforwardly obtained by these methods.

\subsection{Toy simulation and modeling}\label{sec:toys}

Motivated by what we call the ``371-roll scenario''%
  \footnote{More-or-less the most stable pointing option for the \Kepler~Field according to
    the \textit{Explanatory Appendix to the Kepler Project
      call for white papers:\ Kepler 2-Wheel Pointing Control}
    (\url{http://keplergo.arc.nasa.gov/docs/Kepler-2-Wheel-pointing-performance.pdf}).}
  we produced a series of toy (that is, unrealistic in various ways)
  simulated 30-minute coadd images with $\sim$ half-pixel drift and 1-arcsec jitter.
Drift and jitter were simulated by moving the boresight of the simulated telescope
  every 6~seconds during the integration.
An aperture is examined around a single target,
  where the aperture is large enough to contain the full drift path
  of the target over the 24-hour observing window.
(This may be an over-optimistic assessment of what can be downlinked.)
The input (true) positions and brightnesses of the stars in the simulation are
  taken from the \Kepler\ Input Catalog.
The input (true) PSF is a very much simplified double-Gaussian.
Variations are introduced in the (true, toy) flat-field,
   drawn pixelwise iid from a Gaussian with 1~percent standard deviation.
We produced a series of 48 simulated coadds,
   corresponding to one day's observations between rolls.

\begin{figure}
\includegraphics[width=0.49\textwidth]{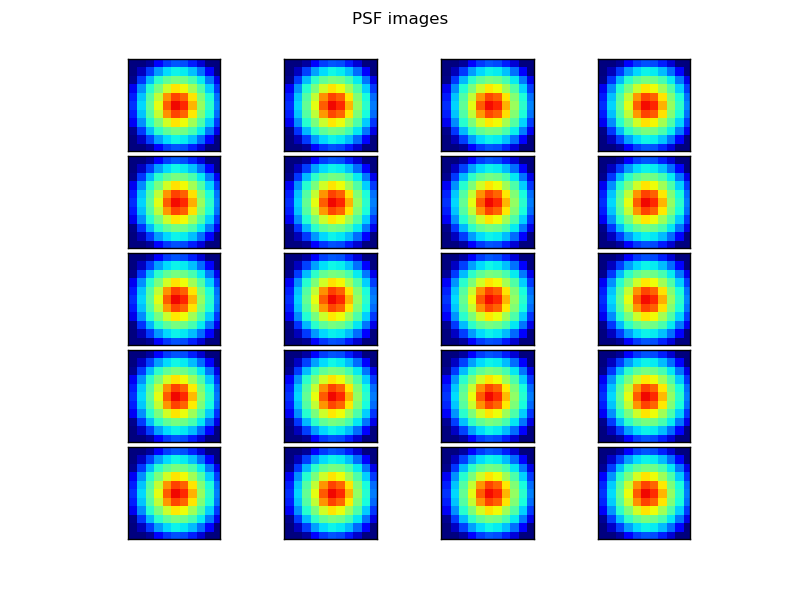}
\includegraphics[width=0.49\textwidth]{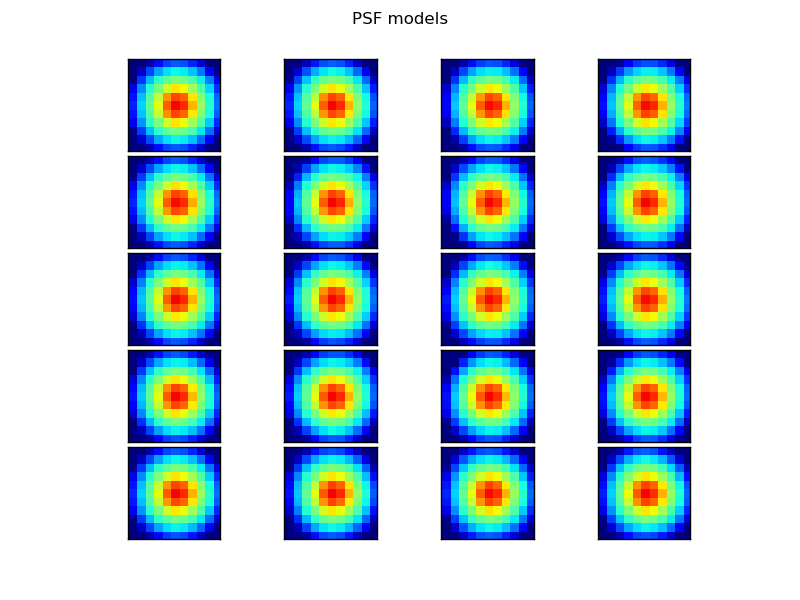}
\includegraphics[width=0.49\textwidth]{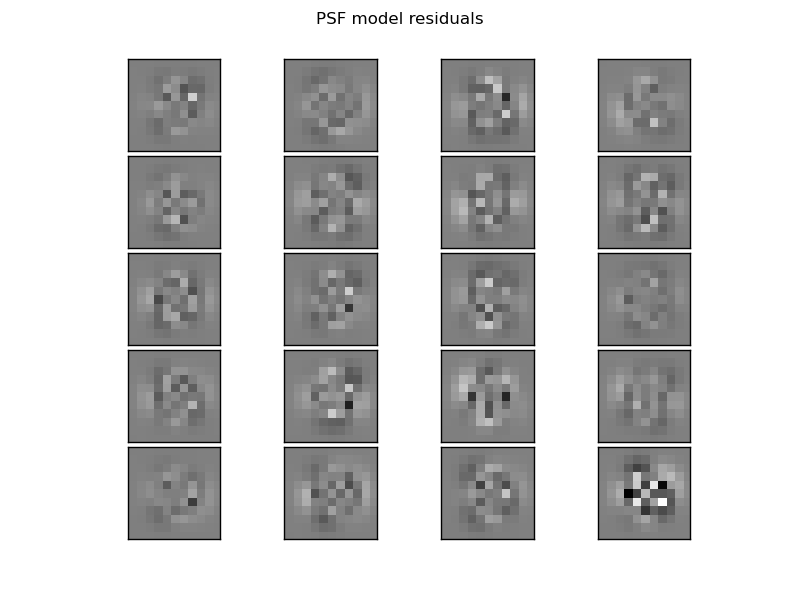}
\caption{Toy simulated two-wheel \Kepler\ PSF-determination data and models.
\textsl{(left)}~``Perfect'' (that is, extremely high signal-to-noise) PSF simulated image, shown on a log scale with a dynamic range of $10^6$.  These simulations were made with a simple PSF model but then 371-roll-level drift and jitter in s/c pointing.
\textsl{(middle)}~Model maximum-likelihood PSFs.  These are 4-component Gaussian mixtures (23 parameters total); that is, the model used to fit the PSF lives in a different space than the model used to generate the toy data.
\textsl{(right)}~PSF-model residuals.  The stretch is $\pm 10^{-4}$.\label{fig:toypsf}}
\end{figure}

In order to simplify the modeling task,
  in addition to the science exposure
  we rendered a ``perfect'', bright, isolated, noise-free star with an unperturbed flat-field.
The star is drifted and jittered in each coadd identically with the full science field.
This ideal star is used to independently learn (fit, estimate) the PSF model in each coadd.
For a PSF model we use a very simple mixture of 4 Gaussians.
That is \emph{we used a different model to generate the data than we used to fit the data},
  which is conservative and realistic in this context.
The fake PSF-determination data and a model of it are visualized in \figurename~\ref{fig:toypsf}.

\begin{figure}
\includegraphics[trim=0.75in 0.5in 0.75in 0.5in, width=0.49\textwidth]{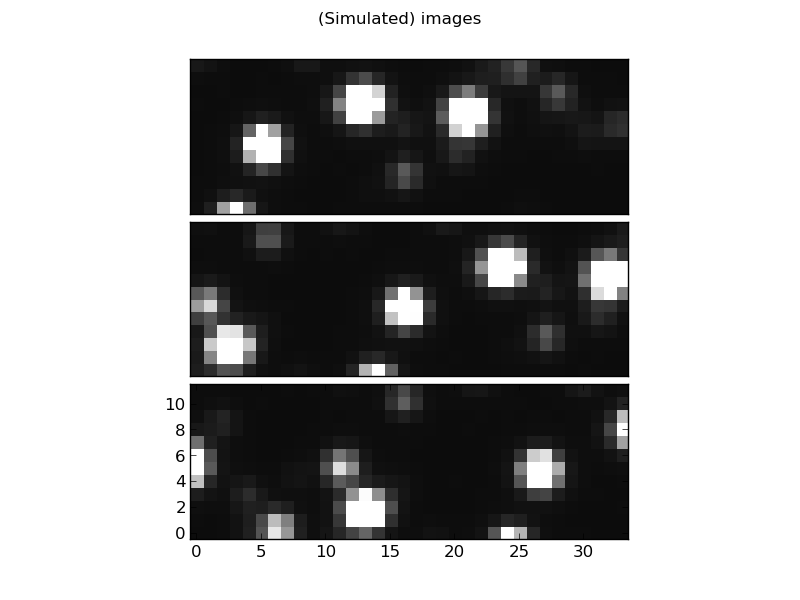}%
\includegraphics[trim=0.75in 0.5in 0.75in 0.5in, width=0.49\textwidth]{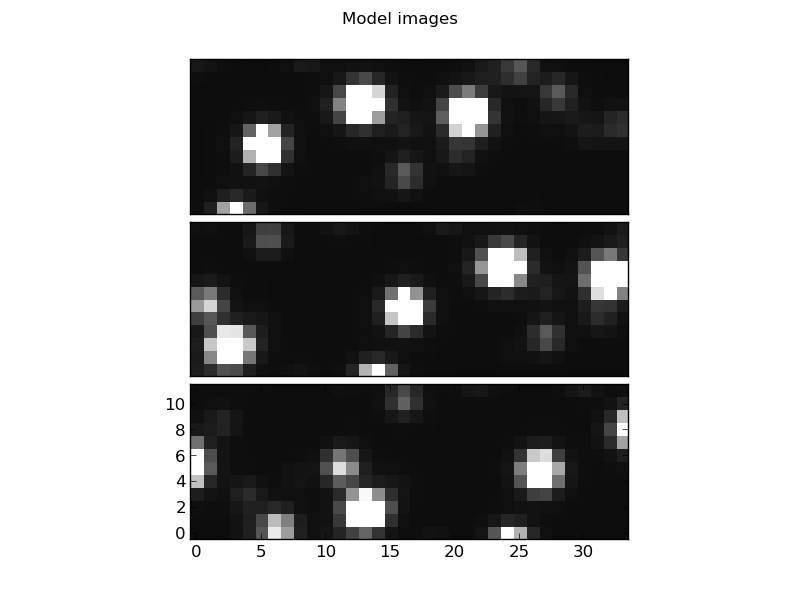}
\caption{Toy simulated two-wheel simulated science data \textsl{(left)} and models \textsl{(right)}.
The model images are created by fitting the amplitude of a (rigidly fit) PSF model for each image,
  and thereby inferring the brightness of each source.
The three images shown are the first, middle, and last images
  in a 48-image sequence in the 371-roll scenario.\label{fig:toydata}}
\end{figure}

After learning a PSF model for each image,
  we infer the brightness of each source in the field.
It is assumed that we have the true source positions from the \Kepler\ Input Catalog
  (the same positions used to produce the simulations, that is, truth),
  and the multi-Gaussian model of the drift-path-wrecked PSF.
This model is unrealistic because we have only considered a single PSF and only two-dimensional
  field shifts in generating the PSF;
  in reality there would be a model of the PSF over the field
  and three degrees of freedom to the s/c drift model (pointing and orientation).
The toy data and models are visualized in  \figurename~\ref{fig:toydata}.

\begin{figure}
\includegraphics[width=0.85\textwidth]{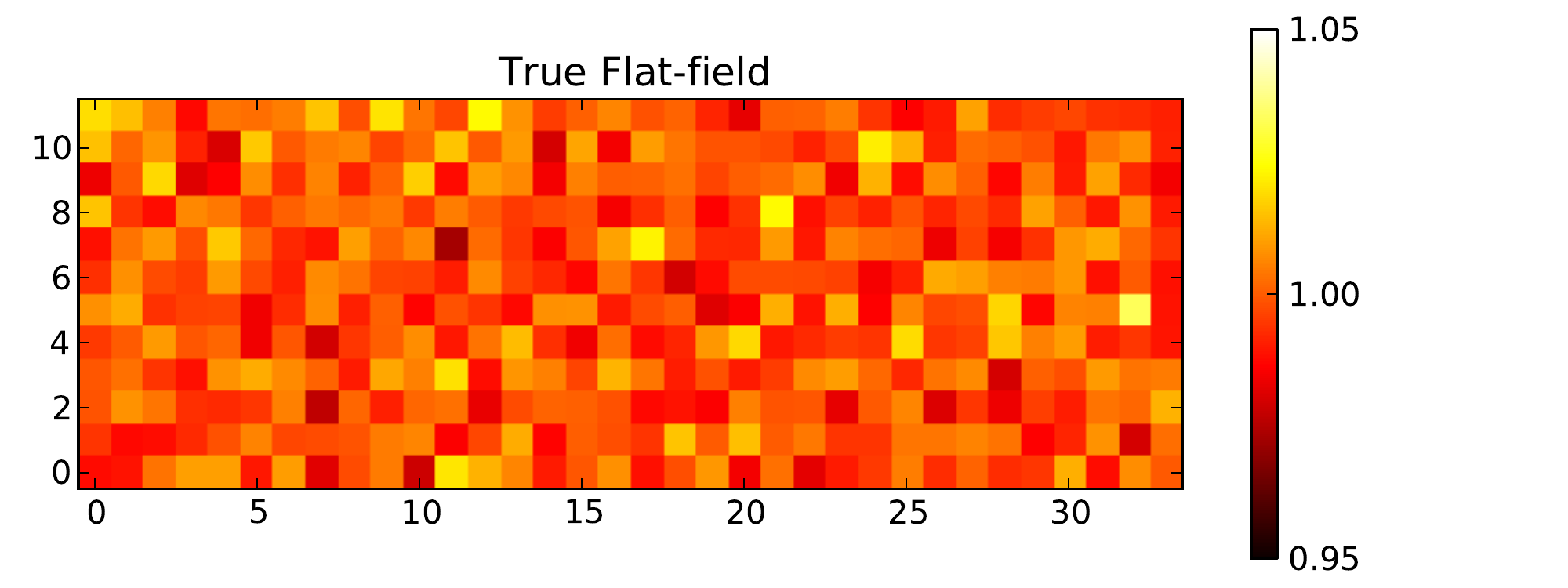}
\includegraphics[width=0.85\textwidth]{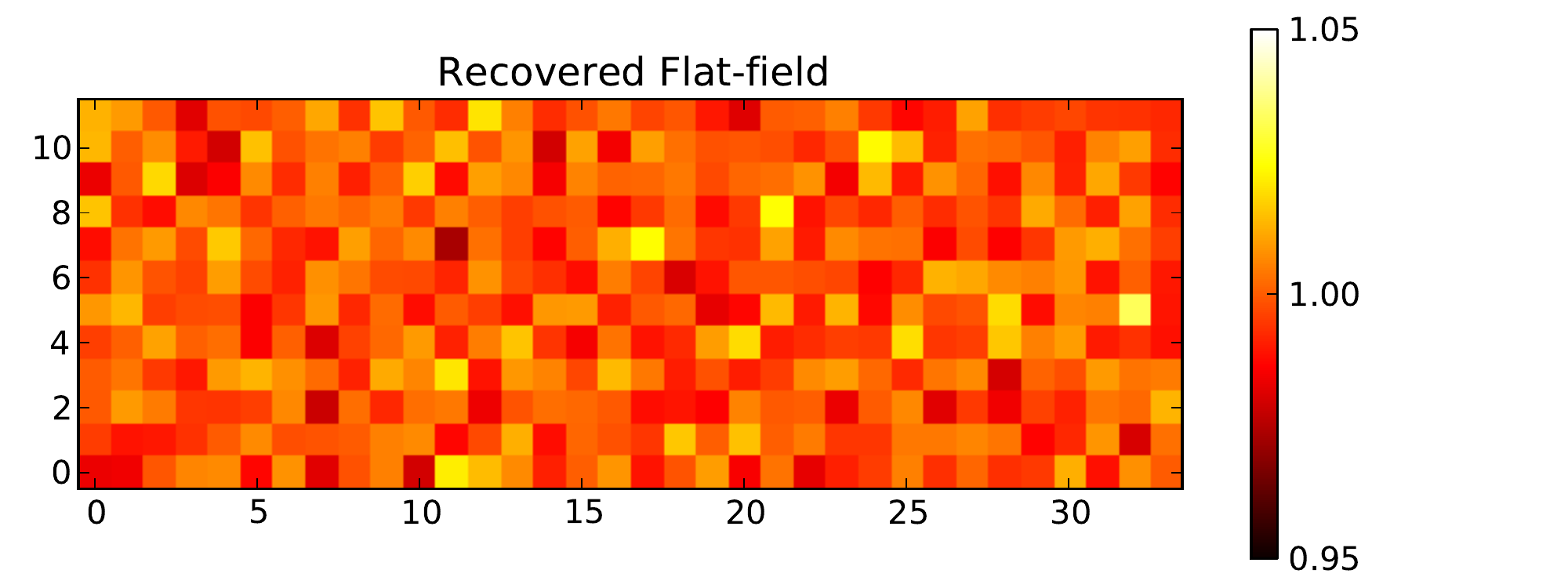}
\includegraphics[width=0.85\textwidth]{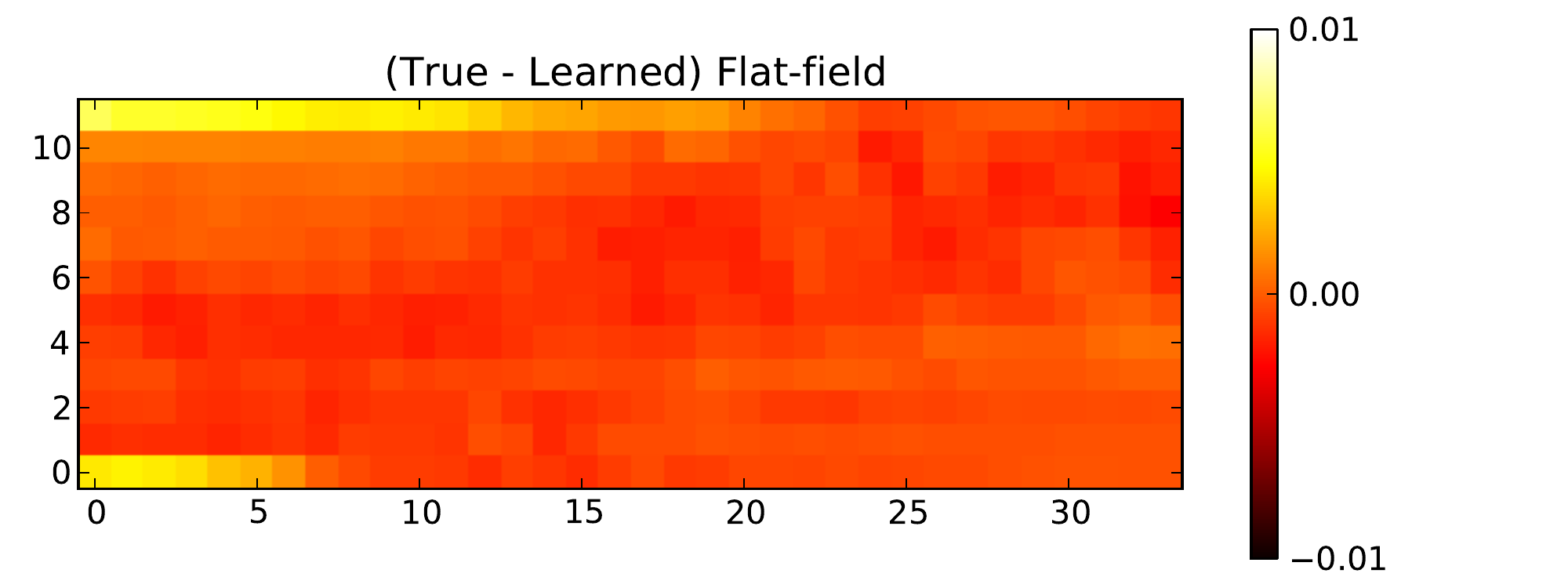}
\caption{Toy simulated true flat-field \textsl{(top left)} and point-estimation thereof \textsl{(top right)},
  found by taking ratios of models to data after fitting the drifted, effective PSF and brightness of every source in the field.
  The true flat-field in the image has Gaussian variation with 1~percent standard deviation.
  We estimated the flat-field with the pixel-wise ratio between the model image and data.\label{fig:toyflat}}
\end{figure}

After fitting for the brightness of each source,
   we estimate the flat-field.
This is estimated simply by taking
  the geometric mean (over the 48 images) of the ratio of the model to simulated images.
That is, we find a pixelwise flat-field that will make our pixel-by-pixel
models best match the observations.
(This method is overly naive; we could do better in principle with simultaneous modeling.)
The true and recovered flat-fields are shown in \figurename~\ref{fig:toyflat}.
We do a good job of recovering the flat-field.

\subsection{Inferring an intrapixel flat-field}\label{sec:intrapixel}

Next, we tested whether there is any chance within these toy simulations
  that we could recover a sub-pixel sensitivity map;
  that is, is it possible to constrain the flat-field at an angular resolution
  higher than the read-out data?
The simulations are similar to the above,
  except this time the simulated images were produced on a pixel grid of twice the true resolution,
  and we randomly generated a flat-field with this same resolution.
After scaling the double-resolution by the double-resolution flat-field,
  the images were binned down (summed) $2\times 2$ to produce images at the detector read-out resolution,
  but now with sub-pixel variations in the sensitivity.
In this simple version then, each pixel has 4 ``zones'' of different sensitivity.

Using our modeling framework,
  it is straightforward to infer such a double-resolution flat-field
  given the normal-resolution images.
As when producing the simulations,
  the model images are rendered at twice the resolution and scaled by a model (parameterized) flat-field
  (that has four times as many parameters as there are device pixels)
  before binning and comparing with the observed images.
As in the previous experiment,
  we first solve for a PSF model,
  then solve for the brightness of each source,
  then solve for the (double-resolution) flat field.
That is, we stay in the fit--fit--fit mode rather than performing simultaneous inference
  of the three components, which should in the end perform better.
In this version, the best-fit flat-field is determined by iteratively solving a linearized least-squares problem:
Starting at a constant flat-field,
  changes to the flat-field pixels are computed that will minimize the difference
  between the model (predicted) images and the observed images.
As before, we fit simultaneously a stack of 48 images, each a 30-minute coadd.
In this experiment, the stars drift horizontally across the field,
  and some regions do not see many stars during the exposure.
To avoid numerical overfitting, we added a prior on the flat-field values,
  with the same variance as the distribution from which the true flat-field was drawn ($\sigma = 2$~percent).

\begin{figure}
\includegraphics[trim=0.75in 0.25in 0.75in 0.25in, width=0.49\textwidth]{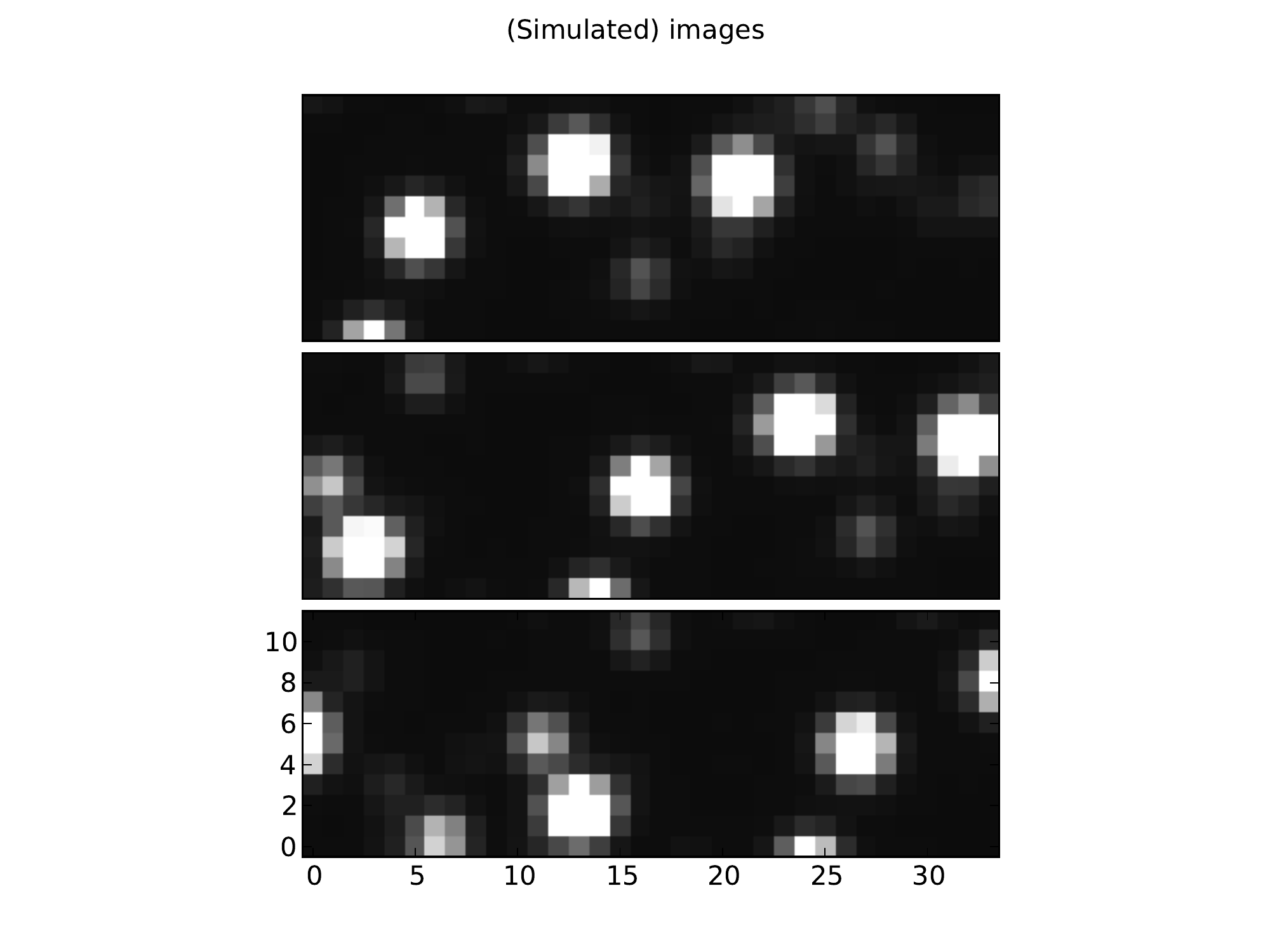}%
\includegraphics[trim=0.75in 0.00in 0.75in 0.00in, width=0.49\textwidth]{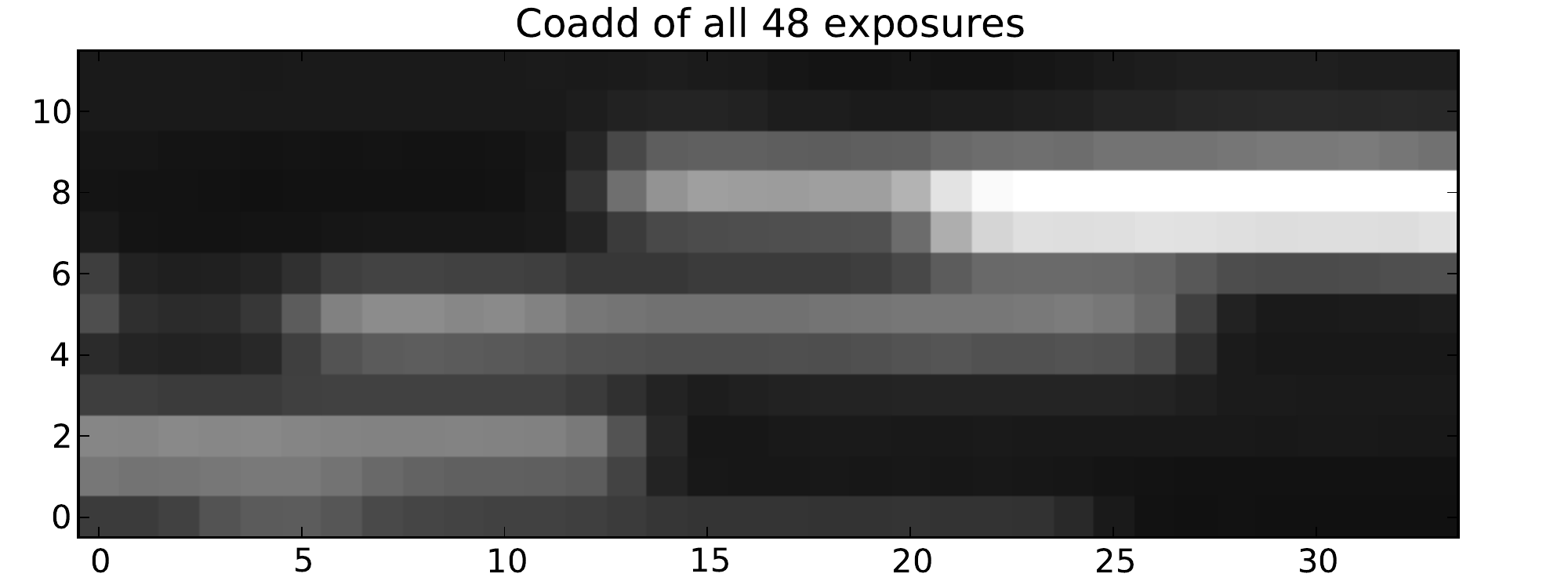}
\caption{Three toy images from the simulated data with $2\times 2$ sub-pixel 2-percent flat-field variations,
  and the coadd of all 48 toy images, showing the 371-roll drift and the coverage of this field patch by stars.
\label{fig:intra-data}}
\end{figure}

\begin{figure}
\includegraphics[width=0.85\textwidth]{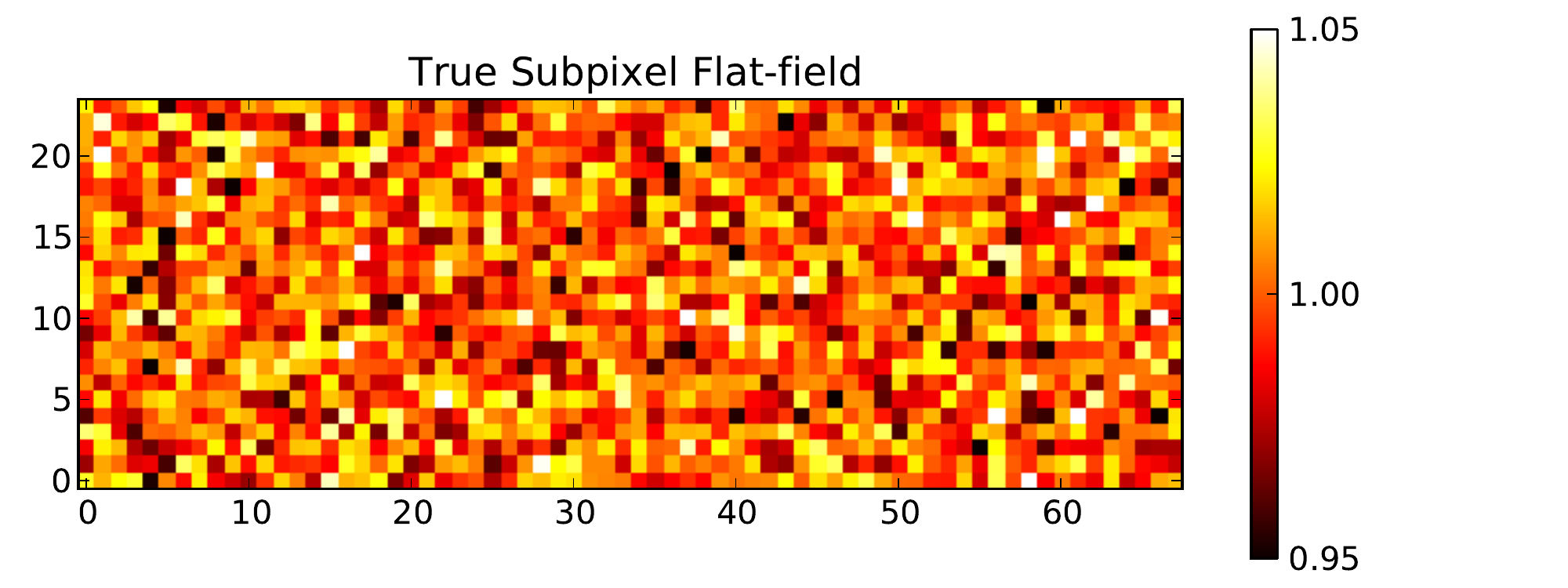}
\includegraphics[width=0.85\textwidth]{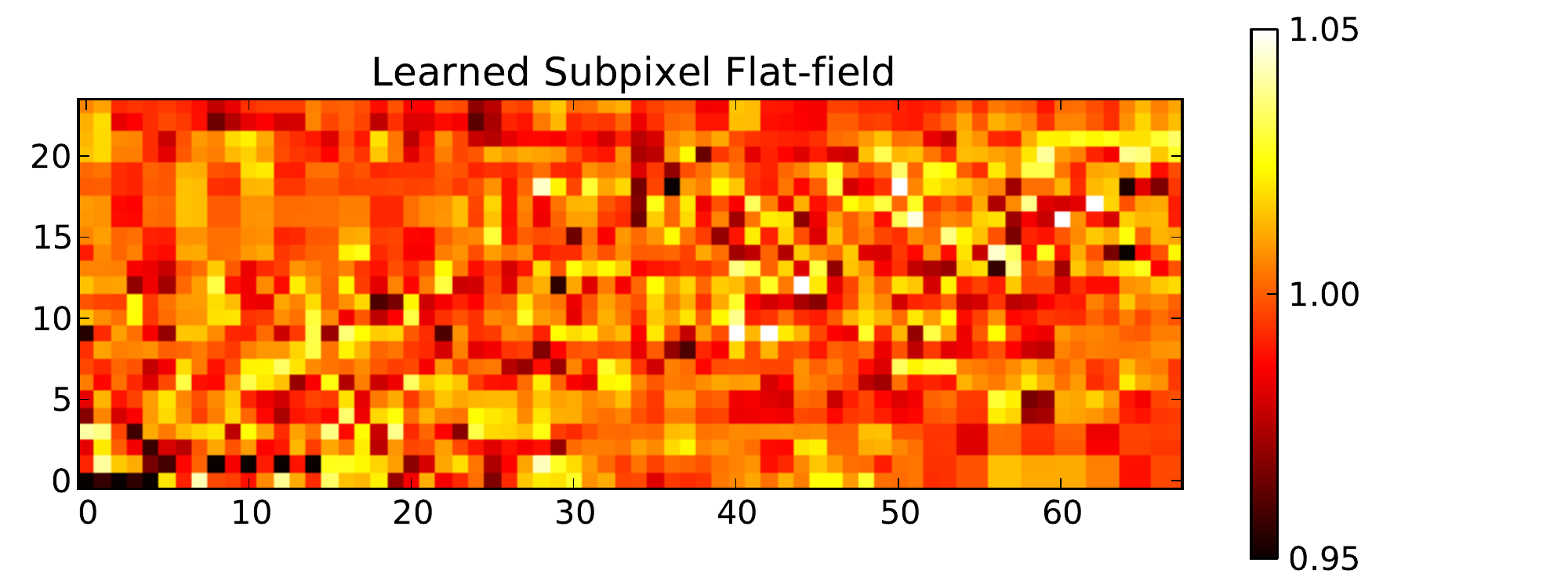}
\includegraphics[width=0.85\textwidth]{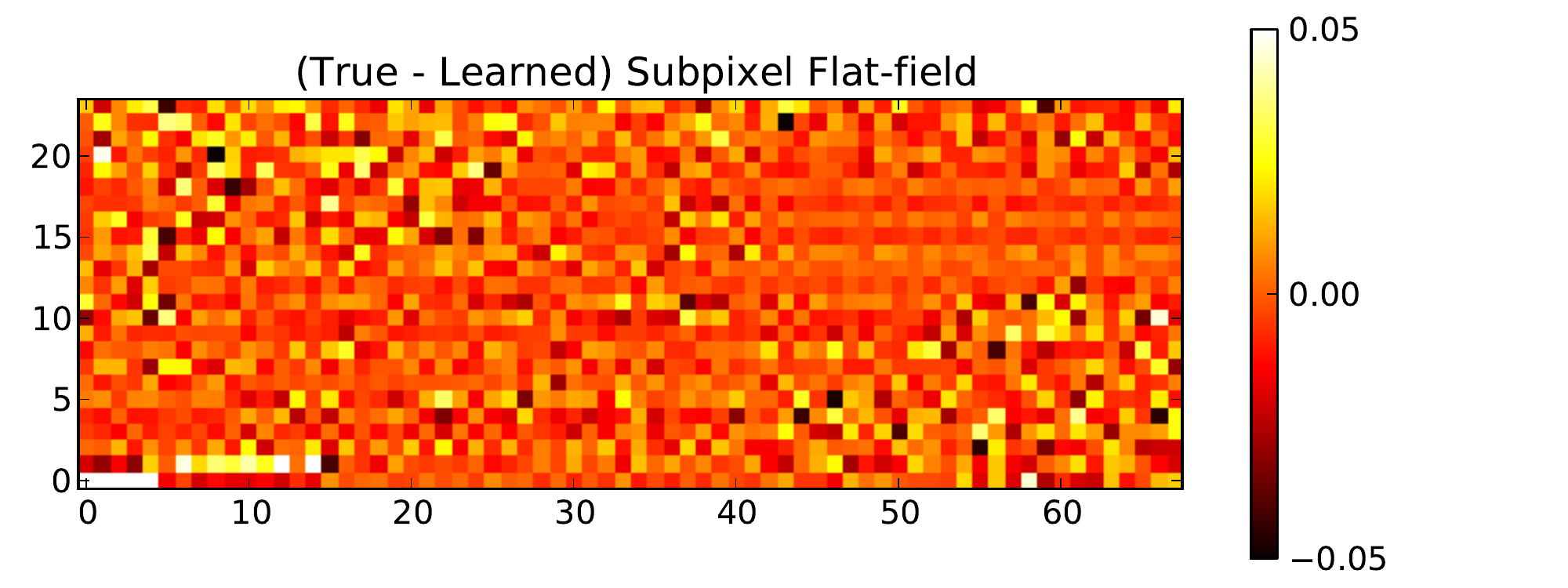}
\caption{Demonstration that we can fit the intrapixel flat---a flat-field with multiple parameters per pixel---at least in principle.
True sub-pixel flat-field, estimated sub-pixel flat-field, and residuals.
  These objects have four times as many parameters as pixels in each of the 48 individual simulated data read-outs.
  Because of the regularization we added to avoid overfitting,
  regions that did not see a fairly bright star are relatively poorly constrained
  and therefore tend to keep the flat-field near unity, and tend to fit original-sized pixel blocks.
Regions through which bright stars drift are much better constained and very closely approximate the true flat field.
In practice, we would likely fit much larger series of images (months or more),
rather than the 48-exposure series used here, so performance could in principle become excellent.\label{fig:intrapixel}}
\end{figure}

The results---shown in \figurename s~\ref{fig:intra-data} and \ref{fig:intrapixel}---%
  are that we can successfully recover a sub-pixel flat-field (sensitivity map),
  especially in regions that are jointly constrained by having stars drift through.
In other words, the quality of the intrapixel flat recovery is a strong function of the history
  of detector illumination, but it is better where we need it to be better (on the stars).
This experiment used only 48 exposures,
  while in practice we would likely attempt to infer a finer sensitivity map
  over a much larger number of exposures: likely months or more of observations.
The degraded pointing and drift of \Kepler\ in two-wheel mode is actually a boon here,
  since it increases the diversity of stars that touch different combinations of pixels.

\textbf{These experiments justify great optimism about the prospects for modeling the \Kepler\ data in the two-wheel era.}
They also show that the drift is not just a curse but also a blessing:
The movement of target stars over the device
  permits determination of very detailed instrument calibration information.

\subsection{Photometry and modeling}\label{sec:photometry}

Here we demonstrate in an extreme toy example that precise image modeling can
facilitate and improve photometry. To this end we first simulate
high-resolution PSFs for both \Kepler's nominal and two-wheel operation
era. As in the experiment above, we use a simplified two component
Gaussian PSF model and assume jitter, normally distributed with a
standard deviation of 1~arcsec. While in the previous experiment
we assumed a minimal drift in the ``371-roll scenario'', here we work
with a much larger drift of 1.4~deg in 4 days as described in the
call\footnote{\url{http://keplergo.arc.nasa.gov/docs/Kepler-2wheels-call-1.pdf}},
which adds up to a total drift length of 27~arcsec for a 30~min
coadd.  This extreme example is designed to maximize the observable impact of modeling.

\begin{figure}[t]
\begin{center}
\footnotesize
\hspace{-2mm}Hi-res PSF \hspace{18mm} Simulated image \hspace{16mm} Model image
\hspace{18mm} Difference\\
\rotatebox{90}{\hspace{6mm}Drift case \hspace{6mm}Nominal operation }\includegraphics[width=0.95\textwidth]{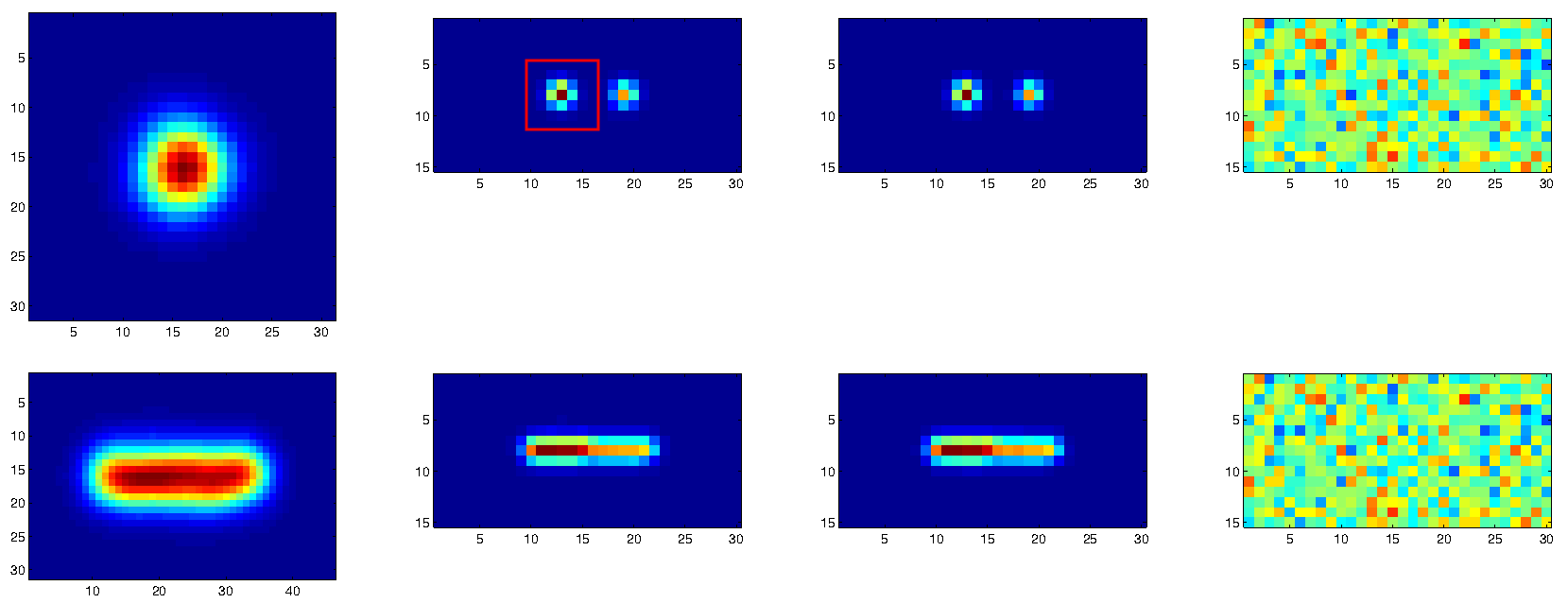}
\caption{Two-star toy example to demonstrate that precise image
  modeling can facilitate and improve photometry over aperture
  photometry. The stars are approximately 25~arcsec apart. 
  From \emph{left} to \emph{right}: high-resolution PSFs
  ($4\times$ higher-resolved, that is, at a resolution of 0.995~arcsec),
  noisy coadds, model and difference images (at \Kepler's nominal
  resolution of 3.98~arcsec). In the simulation and model images
  the dynamic stretch is $10^6$, in the difference image $5*10^2$. The
  \emph{top} row shows the case for nominal operation with functional
  reaction wheels, the \emph{bottom} row shows the case in the
  two-wheel era with an evident drift of 27~arcsec for a 30~min exposure in the worst case scenario (cp.
  \url{http://keplergo.arc.nasa.gov/docs/Kepler-2wheels-call-1.pdf}). The
  red frame depicted in the observed image in the top panel shows the
  optimal (in the sense of least error) aperture used for aperture
  photometry.
  \label{fig:photometry}}
\end{center}
\end{figure}

Again, drift and jitter were simulated by moving the boresight
of the simulated telescope every 6 seconds during a total integration
time of 30 minutes. The left panel of \figurename~\ref{fig:photometry} shows
4x higher-resolved images of the generated PSFs at a resolution of
0.995~arcsec. The panel right next to it shows the convolved and
noisy simulated images of two stars with variable brightness,
approximately 25~arcsec apart. The resolution is at \Kepler's
native pixel scale of 3.98~arcsec. Noise was simulated by drawing
iid from a Gaussian with a standard deviation of 150 e$^-$ assuming a
dynamic range of $1.1\times 10^6$ e$^{-}$ \cite{gilliland2011}. For model
fitting we assume knowledge of both the star positions (for real data
provided by \emph{Kepler's Input Catalog}) and PSF. The third and
fourth columns of \figurename~\ref{fig:photometry} show the model images and
their corresponding differences when compared to the simulated
images.

Table~\ref{tab:photometry} summarizes the photometric results
of our model fitting approach and compares them with the results of
optimal aperture averaging. The values reported in
Table~\ref{tab:photometry} have been averaged over 100 noise
realisations and suggest that image modeling does improve over simple
aperture photometry even in \Kepler's two-wheel era. Note that the
footprints of stars separated less than the drift length within a
coadd, will overlap in general (which is the case in our toy
example). In this case it is no longer possible to define an aperture
that would allow for high-precision photometry, such that aperture
photometry won't work for nearby stars (with a distance less than the
drift length per coadd). In contrast, our modeling approach is still
able to yield photometric information with a higher accuracy than
aperture photometry in the nominal case! The slight loss in accuracy
and precision suggests that a smaller drift is in general preferable,
which could be accomplished by shorter exposure times. However, there
is tradeoff to be explored between smaller drift and lower SNR through
shorter exposure times (also see our discussion on exposure time in
Section \ref{sec:operations}).

\subsection{PSF modeling}\label{sec:psf}

We have shown (above) that fitting images with a good PSF provides better
  photometry than aperture photometry.
What we can't show just yet is that the PSF will be comprehensible or known
  at the requisite precision.
This represents a very important challenge for the data-processing side of two-wheel \Kepler\ operations.

Drift will cause the PSF to vary from exposure to exposure due to the
random nature of the jitter introduced by the pointing inaccuracy of
the spacecraft. While previously, due to stable pointing, the jitter
contributed to the \Kepler's optical PSF in form of an enlarged PSF, in
the two-wheel era jitter will render the PSF to change from exposure
to exposure. Therefore, we believe that precise PSF modeling is required
on an individual exposure level to maintain high-precision photometry.

The PSF in two-wheel mode can be decomposed in two distinct
components: one that describes \Kepler's (unchanged) system PSF and a
second component that describes the global motion/drift of the
spacecraft during exposure.

For modeling the former, previous \Kepler\ data is invaluable, which
will tie up on the work of \cite{bryson2010} and allow a precise model
of \Kepler's instrument PSF with the help of a detailed analysis of the
temporal stability of \Kepler's pixel response function (PRF). While in
\cite{bryson2010} only isolated non-saturated star images recorded
during \Kepler's commissioning phase have been used to build a model of
Kepler's PRF as a piecewise-continuous polynomial on a sub-pixel mesh,
in principle any non-saturated star images from the \Kepler\ Input
Catalog can be used. In general more data will facilitate the
estimation and at the same time improve modeling accuracy as
demonstrated in \cite{magain2007}.

The second PSF component is stemming from spacecraft motion/drift and
is affecting the image in a globally consistent way. Hence, with
precise knowledge of the detector's geometry and its possibly time and
space-varying PRF, all star images from an individual exposure will
help to constrain this (to some extent unkown) motion. Any additional
knowledge about the spacecraft's pointing position will facilitate the
estimation process. Note, that for an individual exposure we can be
agnostic about the chronology and temporal evolution, as we are
interested in its time integrated effect on the image only. The
incorporation of a geometric model allowing for physically plausible
motion only has been applied with great success to the problem of
camera shake removal with non-stationary blur within the computational
photography and computer vision community recently
(\citealt{fergus2006,whyte2010,hirsch2011,koehler2012}).  


Although from an image-processing perspective all necessary tools are
readily available and have proven extremely successful in a number of
non-trivial imaging applications recently, PSF modeling with
high-precision photometric accuracy is a non-trivial task.  At the
same time this is a promising and exciting direction to be explored for
existing \Kepler\ data and two-wheel data in the future.


\section{Changes to operations}\label{sec:operations}

Here are some thoughts about operational changes, relevant to image
modeling.

\paragraph{exposure times:}
In the pointing-stability best-case 371-roll scenario,
  the long cadence of 30-min exposures can be continued without change.\footnote{We note also that the 371-roll scenario maintains the s/c in a very stable thermal environment, which may simplify many aspects of calibration.}
If the science program obviates the 371-roll,
  then exposure times might have to get shorter.
In a worst-case scenario%
  ---of either a failure of the 371-roll predictions,
  or else requirements that make it impossible to implement---%
  we advise setting exposure times to obtain something close to 1 pixel
  of jitter or drift during the exposure.
Much less than this, and the exposures are ``over-resolving'' the jitter
  (thereby wasting telemetry bandwidth)
Much more than this, and the angular resolution and PSF estimation is
  likely to degrade in ways that affect photometric precision.
One interesting consequence of sticking with the \Kepler~Field and shortening
  exposure times is that the mission could deliver better information
  about transit-timing variations, transit-duration variations, and asteroseismology.

\paragraph{telemetry of pixels:}
Even in the 371-roll scenario, stars drift steadily across the focal plane,
  requiring either much larger apertures to be downlinked,
  or else adaptive aperture control from exposure to exposure.\footnote{All scenarios considered in the \textit{Explanatory Appendix} assume that the two remaining wheels will be used to correct disturbance torques about the Z and Y axes.  The wheel orientations are such that any Y-correction will induce a roll about the X axis; optimal scenarios minimize Y disturbance torques.  Boresight roll will result in variable drift across the detector, but has minimal impact on the analysis described above.}
The former comes at telemetry bandwidth cost,
  the latter at satellite table or software upgrade cost.
From an image-modeling perspective, we don't have a position on this;
  if anything the larger apertures are good,
  because more pixels equals more calibration information.
From a scientific perspective, we much prefer a change to flight tables or software;
  bandwidth is valuable.

\paragraph{target list:}
Exposure time shortening or aperture enlarging will require more telemetry bandwidth per target star.
If necessary, we recommend shrinking the target list around two parent populations:
One is M stars, where small, habitable-zone planets seem to be abundant.
The other is G stars, where only true Earth-analogs can live (and true Jupiter analogs).
Some of these issues are addressed in greater detail in our companion \documentname\ (Montet \etal).
In principle a lot is now known about the detectability of interesting planets
  as a function of stellar variability patterns.
This knowledge is also valuable for shrinking target lists.

\paragraph{telemetry of housekeeping data:}
Because recovery of pointing history is so important to image modeling,
  it makes sense to telemeter down housekeeping data from the
  satellite fine guidance sensors and tracking and pointing software.
We suspect that even maximal downlink of pointing-related housekeeping data
  (s/c thermal and star tracker HK, the latter ideally at 5~Hz)
  is tiny compared to the scientific data.
If so---and if it is not already being downlinked---%
  we would recommend making the relevant changes to subscribe to these HK data.

\paragraph{calibration observations:}
The image modeling in our demonstration projects (above) is successful
  because there are data spanning a wide range of conditions
  (temperature conditions in the data-driven modeling example
   and pointing conditions in the toy simulations).
There is no reason to stop at the data diversity delivered by the
  stochastic forces of Solar insolation.
In specified calibration periods%
  ---or more radically throughout the two-wheel mission---%
  it makes sense to pull the focus (if possible),
  dither the pointing deliberately (if software permits),
  and exercise the exposure time control.
These changes will give vital calibration information about
  the PSF, the flat-field, and the DC signals.
In the end, the self-calibration provided by modeling the science
  data might be sufficient.
But if or where it isn't, a small amount of calibration data can be enormously valuable.

\section{Acknowledgments}

It is a pleasure to thank Sameer Agarwal (Google) and Keir Mierle
  (Locu) for releasing the \project{Ceres} non-linear least squares
  solver (\url{https://code.google.com/p/ceres-solver/}) and for making
  suggestions that enabled this work.
We benefitted from useful conversations with and assistance from
  Ross Fadely (NYU),
  Malte Kuhlmann (MPI-IS),
  Hans-Walter Rix (MPIA), and
  Ben Weiner (Arizona).
DWH, RF, and DFM are all partially supported by NSF grant IIS-1124794.
MH acknowledges support from the European Research Council in the form of a Starting Grant with number 240672.
BTM is supported by the NSF Graduate Research Fellowship grant DGE-1144469.


\clearpage

\clearpage

\begin{table}[b]
\begin{center}
\begin{tabular}{c|c}
\emph{Nominal operation} & Relative error ($\times 10^{-4}$)\\ \hline
Aperture averaging &  $21.50 \pm 10.35$ \\ 
Modeling approach  &  $5.55 \pm 3.00$ \\[5mm] 
\emph{Two-wheel operation}  & \\ \hline
Modeling approach &  $8.21 \pm 4.46$ \\
\end{tabular}
\end{center}

\caption{Photometric results for the toy example shown above
  (\figurename~\ref{fig:photometry}). The PSF modeling approach improves
  photometric accuracy over aperture averaging even in the two-wheel
  era with a drift as large as 27~arcsec for a 30~min exposure.
  For averaging optimal apertures have been used (exemplary one is depicted as red
  frame in the top panel in the \figurename\ above). Note, that in the drift case the footprints of the two
  stars overlap, such that photometric measurement is not possible by
  simple aperture photometry.}

\label{tab:photometry}
\end{table}

\end{document}